\documentclass[12pt,a4paper]{article}
\usepackage{amsmath}
\usepackage[dvips]{graphicx}
\usepackage{times}
\begin{document}
\begin{titlepage}
\title{On the double-ridge  effect at the LHC }
\author{S.M. Troshin, N.E. Tyurin\\[1ex]
\small  \it Institute for High Energy Physics,\\
\small  \it Protvino, Moscow Region, 142281, Russia}
\normalsize
\date{}
\maketitle

\begin{abstract}
 We  discuss a possible explanation of the double-ridge observation in pPb--collisions at the LHC
 emphasizing that
this double structure in the two-particle correlation function can result from 
the rotation of the transient state of matter. 
\end{abstract}
\end{titlepage}
\setcounter{page}{2}

The ridge-like  structure in correlation function of two particles has been  observed for the first time  at RHIC in the peripheral collisions 
of nuclei  in the near-side
jet production (cf. recent paper \cite{rhicr} and references therein for the earlier ones). It was observed  that the secondary particles have
 a narrow $\Delta\phi$ two-particle correlation distribution (where $\phi$ 
is an azimuthal angle) and wide distribution over $\Delta\eta$  ($\eta$ is a pseudorapidity). This phenomenon was called ridge effect and associated
usually with the collective properties of a medium produced under interaction. 

For the first time, a similar effect was revealed by
the CMS Collaboration \cite{ridgecms} in pp--collisions and it becomes clear that a
deconfined phase formation has the qualitative similarity  in pp- and AA-collisions. This feature has
been confirmed by the observation of the ridge effect in PbPb-collisions by  ALICE, ATLAS and CMS \cite{al,at,cm}.
We emphasize
further that significant role belongs to peripheral  collision dynamics controlled in AA--collisions by the experimental 
conditions and in pp--collisions by the transition into the reflective scattering mode \cite{intja} at the LHC energies.
Of course, there should be a quantitative difference between pp- and AA-collisions and the signal of ridge should be more prominent in AA-interactions.

The peripheral interactions and quark-pion liquid formation in the transient state of interaction \cite{multrev} would lead to the coherent rotation
which can provide explanation for the double-ridge effect observed recently by the ALICE and ATLAS collaboration in pPb-interactions at
$\sqrt{s_{NN}}=5.02$ TeV \cite{alpb,atpb}.  Those most recent experimental data probe net effects related to the matter formed under interaction
since they were
obtained by subtraction of the low multiplicity events (ALICE) or  contributions from recoiling dijets (ATLAS). Thus, they provide
quantitative estimates  of the collective effects contribution.

It should be noted that peripheral mechanism of pPb-interactions  is controlled dynamically
by the reflective scattering mode which is prominent at the LHC energies \cite{intja}. 
The details will be described further.
Now we briefly repeat the main points of the 
production mechanism  proposed in \cite{intje}. It is based on the
geometry  of the overlap region and dynamical properties of
the transient state in hadron interaction\footnote{Of course, this is not unique mechanism leading to the appearance of the ridge, 
description of the other ones can be found
in the recent review paper \cite{rev}}. The above picture assumes  that the deconfinement takes place at the initial stage of interaction.
The transient state of matter appears then as a rotating medium of massive quarks and pions which hadronize and 
form multiparticle state at the final stage of interaction.  The essential point which is needed  for the presence of  this
rotation is a non-zero impact parameter in the collision. We believe that this qualitative picture seems applicable 
to pp, pA and AA-interactions. However, the peripheral mechanism of pp-interaction (and of pA-interactions also), 
as it was already mentioned, is controlled dynamically
by the transition to the reflective scattering mode at the LHC energies. 
Due to the reflective scattering \cite{intja} the inelastic overlap function $h_{inel}(s,b)$,
\[
h_{inel}(s,b)\equiv\frac{1}{4\pi}\frac{d\sigma_{inel}}{db^2},
\]
has a peripheral impact parameter dependence in the region of $\sqrt{s}>2$ TeV  \cite{intja}.
Unitarity equation 
for the elastic amplitude $f(s,b)$ in the impact parameter representation rewritten at high energies  has the simple form
\begin{equation}
 \mbox{Im} f(s,b)=h_{el}(s,b)+ h_{inel}(s,b)
\end{equation}
and $h_{inel}(s,b)$ is the sum of all inelastic channel contributions to the unitarity equation.
Consider for ilustration the case of pure imaginary elastic scattering amplitude, i.e. $f(s,b)\to if(s,b)$.
Then the inelastic overlap function is related to the elastic scattering amplitude by the simple relation 
\[
h_{inel}(s,b)=f(s,b)(1-f(s,b)).
\]
Reflective scattering mode implies the saturation of the unitarity limit at small impact parameters, e.g.  $f(s,b=0)\to 1$
at $s\to \infty$. Thus, the peripheral form of the inelastic overlap function is just an effect of the reflective 
scattering\footnote{The elastic scattering amplitude $f(s,b)$ is supposed to be a monotonically decreasing function of the 
impact parameter}.
According  to peripheral impact parameter dependence of $h_{inel}(s,b)$, the mean multiplicity 
\[
 \langle n\rangle (s)=\frac{\int_0^\infty bdb  \langle n\rangle (s,b) h_{inel}(s,b)}
{\int_0^\infty bdb h_{inel}(s,b)}
\]
obtains the maximal input from the collisions  with non-zero impact parameter values.
Thus,  the events with signicant values of multiplicity at the LHC energies would  correspond
to the peripheral hadron collisions \cite{intja}. At such high energies there is a dynamical
selection mechanism of peripheral region in impact parameter space.  In the nuclear collisions
 such selection is provided by the relevant  experimental conditions. 

The  geometrical picture of hadron collision  at non-zero impact parameters
implies \cite{multrev}  that the generated massive
virtual  quarks in the overlap region  could obtain very large initial orbital angular momentum
at high energies (Fig. 1).
\begin{figure}[htb]
\hspace{3cm}
\resizebox{7cm}{!}{\includegraphics*{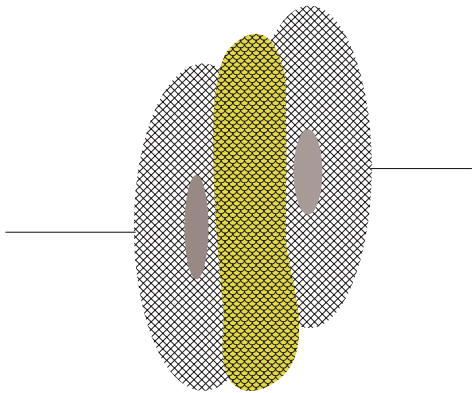}}
 \caption[illyi]{{\small\it Overlap region in the  initial stage of the hadron
 interaction, valence constituent quarks are located in the central part of hadrons}.}
\label{ill5}
\end{figure}

Due to strong interaction
between the  quarks this orbital angular momentum  leads to the coherent rotation
of the quark system located in the overlap region  in the
$xz$-plane (Fig. 2). This rotation is similar to the  rotation of the liquid
where strong correlations between particles momenta exist. 
Thus, the non-zero orbital angular momentum should be realized  as a coherent rotation
of the quark-pion liquid  as a whole. 
The assumed particle production mechanism at moderate transverse
momenta is the simultaneuos excitations of  the  parts of the rotating transient state (of  massive constituent
quarks interacting by pion exchanges) by the valence constituent quarks. We refer here to proton-proton interactions, in pA-interactions
the effects should be enhansed due to many collisions but would remain  qualitatively the same.

Since the transient matter is strongly interacting, the excited parts
should be located closely  to the periphery  of the rotating transient state otherwise absorption
 would not allow quarks and pions  leave the interaction region (quenching). 

The mechanism is sensitive
 to the particular  direction of rotation and to the rotation plane orientatation
leads to the narrow distribution of the two-particle correlations in $\Delta\phi$. However,
these correlation could have a broad distribution versus polar angle ($\Delta\eta$) (Fig. 2).
 Quarks in the exited part of the cloud
could have different values of the two components of the momentum (with its third component
lying in the rotation  $xz$-plane) since the exited region is spatially extended. 
 The rotation with orbital angular momentum $L(s,b)$ contributes to the $x$-component of the transverse 
momentum and does not contribute to the $y$-component of the transverse momentum, i.e.
\begin{equation} \label{ptx}
\Delta p_x=\kappa L(s,b)
\end{equation}
while
\begin{equation} \label{pty}
\Delta p_y=0.
\end{equation}

\begin{figure}[h]
\begin{center}
\resizebox{7cm}{!}{\includegraphics*{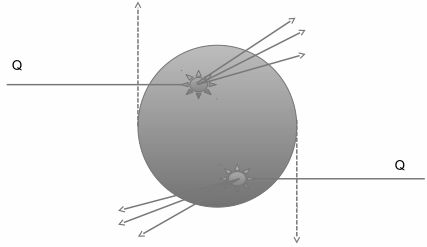}}
\caption{\small \it {Interaction of the valence constituent quarks with rotating quark-pion liquid, dashed lines indicate the clockwise rotation of 
the transient matter.}}
\end{center}
\end{figure}
It should be noted that the ALICE and ATLAS  are observing the double-ridge structure with two maximums
in the two-particle correlation function at $\Delta\phi=0$ and $\Delta\phi=\pi$ after subtraction of the low-multiplicity events or
expected dijet events, respectively. In terms of Fig.2 this observation means the presence of the correlations between particles in the 
upper cluster  ($\Delta\phi=0$) 
and between particles from the upper and lower clusters ($\Delta\phi=\pi$).  Lower cluster of particles is the result of the excitation of the transient
matter by the constituent valence quark from the second colliding hadron. Since both particle clusters originate from the coherent rotating quark-pion
liquid, it is evident that the particles from lower and upper clusters should be correlated and this will produce the second observed ridge.

Thus,  the double-ridge structure observed  in the high multiplicity events by the ALICE and ATLAS 
can serve as an experimental manifestation of the coherent rotation   of the transient state of matter generated during the hadron
collisions. The narrowness of the two-particle correlation distribution in the asimuthal angle is the
distinctive feature of this mechanism which reflects the presesence of the rotation plane in every event of particle production at the LHC. 

Other experimentally
observed effects of this collective effect are described in \cite{multrev}.

It should be noted that we believe that the nature of the state of matter 
revealed at the LHC in proton collisions is the same as the nature of the state revealed
  at RHIC under collisions of nuclei.  But, since the LHC energies are significantly higher that the RHIC
energies, the differences should be expected due to the energy-dependent dynamics related to the peripheral form 
of inelastic collisons generated 
by the reflective scattering mode. There should be no ridge effect in pp- and pA-interations at RHIC, while
the double-ridge effect should be observed in the events of particle production in the peripheral AA-collisions at RHIC 
and in pp-collisions at the LHC
(after subtraction procedure of the low-multiplicity or dijet events being performed). Evidently, the double-ridge effect should
be observed by the CMS experiment also  in pPb-collisions with the use of the similar subtractions.

\end{document}